\documentclass[reprint,amsmath,amssymb,superscriptaddress,prl,showpacs,prl]{revtex4-1}
\usepackage{graphicx}
\usepackage{dcolumn}
\usepackage{bm}
\usepackage{amsmath}
\usepackage{amssymb,amsthm}
\usepackage{color}
\usepackage{epstopdf} 
\usepackage{subfig}

\usepackage{epstopdf}
\usepackage{mathbbol}

\let\originalleft\left
\let\originalright\right
\renewcommand{\left}{\mathopen{}\mathclose\bgroup\originalleft}
\renewcommand{\right}{\aftergroup\egroup\originalright}

\begin{document}
\title{Exploring the diluted ferromagnetic p-spin model with a Cavity Master Equation}

\author{Erik Aurell} 
\email{eaurell@kth.se}
\affiliation{KTH – Royal Institute of Technology, AlbaNova University Center, SE-106 91 Stockholm, Sweden}
\author{Eduardo Dom\'{\i}nguez} 
\affiliation{Group of Complex Systems and Statistical Physics, Department of Theoretical Physics, University of Havana, Cuba}
\author{David Machado} 
\affiliation{Group of Complex Systems and Statistical Physics, Department of Theoretical Physics, University of Havana, Cuba}
\author{Roberto Mulet} 
\email{mulet@fisica.uh.cu}
\affiliation{Group of Complex Systems and Statistical Physics, Department of Theoretical Physics, University of Havana, Cuba}

\date{\today}


\begin{abstract}
  We introduce a new solution to Glauber multi-spin dynamics on random graphs.
  The solution is based on the recently introduced Cavity Master Equation (CME), a time-closure turning
  the in principle exact Dynamic Cavity Method into a practical method of analysis and of fast simulation.
  Running CME \textit{once} is of comparable computational complexity as one Monte Carlo run on the same problem.
  We show that CME correctly models the ferromagnetic $p$-spin Glauber dynamics from high temperatures
  down to and below the spinoidal transition. We also show that CME allows a novel exploration of the
  low-temperature spin-glass phase of the model.
  
\end{abstract}

\maketitle


Nature abounds in systems of interacting units with non-trivial dynamical properties. This
leads to similar questions in condensed-matter physics \cite{Onuki}, systems biology \cite{Nowak}, neuroscience \cite{Amit} and neural networks \cite{Gurney},  but also in many practical applications in computer science \cite{mezard2009information} and engineering \cite{Schiff}.
Not surprisingly the study of such systems has led to the development and the use of similar techniques across different scientific communities. 


A common starting point is a Markovian dynamics in continuous time of $N$ discrete interacting variables  $\underline{ \sigma } =\{\sigma_1, \dots \sigma_N\}$.  Such a system is described by a Master Equation defining the evolution of the probability of the states of the system $P(\underline{\sigma})$, \cite{vanKampen,coolen2005theory,Amit}:

\begin{equation}
  \frac{d P(\underline{\sigma})}{dt} = \sum_{\sigma^{'}} \big[ r(\underline{\sigma^{'}} \rightarrow \underline{\sigma}) P(\underline{\sigma^{'}}) -
   r(\underline{\sigma} \rightarrow \underline{\sigma^{'}}) P(\underline{\sigma}) \big]\label{eq:master-equation}
\end{equation}

\noindent where $r$ are transition probabilities from states.

The  solution of the Master Equation (\ref{eq:master-equation}) is in general a cumbersome task and exact results on both stationary states and transients are limited to some special cases~\cite{Glauber63,mayer2004general,coolen2005theory,Amit}. For fully connected  \cite{laughton1996order} and dilute graphs \cite{Semerjian04} it is possible to resort to hierarchical schemes to derive dynamical equations for the probability of some macroscopic observables. A very general one of this type is the Dynamical Replica Analysis\cite{Hatchett05,mozeika2008dynamical,mozeika2009dynamical}.  With this reduction of the dimensionality  the problem becomes tractable,
but  one looses the detailed information about the microscopic state of the system and the results are not exact for
fully connected~\cite{Nishimori}, nor for the transients in 1D ferromagnetic systems~\cite{DelFerraroAurell}.

 A frequently made approximation reduces  (\ref{eq:master-equation}) to a simpler Master Equation for the probabilities of single spin variables $P(\sigma_i)$. However, this is only valid in mean-field-like models\cite{Amit}, or at very high temperatures. An alternative solution that successfully describes the local dynamics of these systems in a wide range of the parameter space was recently suggested in\cite{Aurell}. The problem is then reduced to the solution of a Master Equation for a conditional probability that is subsequently used to recover the full $P(\sigma_i)$ of equation.


In this work we generalize this last approach to systems with multi-particle interactions and explore for the first time its potential to describe processes both near to equilibrium and deep into a phase with multiple minima. We show also that it can be exploited to obtain the ground-state of models with a complex energy landscape.
To keep the notation simple, and to conveniently explore the features of our solution we study the dilute ferromagnetic $p$-spin model with Glauber dynamics. However, the ideas behind our derivation should be clear enough  to allow the straightforward application of our method to other dynamical rules and other Hamiltonians.


The ferromagnetic $p$-spin model is defined by the Hamiltonian :
\begin{equation}
  H = -\sum_{i_1,i_2, \dots i_p} J_{i_{1}i_{2}i_{3}...i_{p}} \sigma_{i_{1}}\sigma_{i_{2}}\sigma_{i_{3}}...\sigma_{i_{p}}
  \label{eq:Hamiltonian-pspin}
\end{equation}

\noindent where $J_{i_1,i_2,\dots i_p}=1$ for all p-tuplets in the set of hyper-edges of a Bethe lattice and zero otherwise, and $s_i \in \{-1,1\}$ are binary variables.  This model is a natural intermediary between spin glasses and structural glasses. Like the former, it is defined by binary variables and fixed in an infinite lattice. With the latter it shares a crystalline state, the absence of quenched disorder, and a finite range of the interactions. 


A standard  dynamics for this model uses a single spin transition rule between states $ r(\underline{\sigma} \rightarrow 
\underline{\sigma^{'}})= r_i({\sigma_1, \dots \sigma_i, \dots} \rightarrow {\sigma_1, \dots -\sigma_i, \dots })$. In addition, we will consider 
that the transition rate of spin $i$  depends only on the state of the spin and its neighborhood; in this case, all p-tuplets it belongs to.
Let us define $\partial i$ as the set of p-tuplets that include spin $i$ and use $\sigma_{\partial i}$ as a notation for the set of spins in $\partial i$ excluding $\sigma_i$ and denote $\sigma_a$ as the group of spins forming the p-tuplet $a$. The transition rate for spin $i$ will be then, for a Glauber dynamics:$r_i(\sigma_i,\sigma_{\partial i}) = \frac{\alpha}{2}  \left[ 1 - \sigma_{i} \tanh \left( \beta  \sum_{a \in \partial i} J_{a} \sigma_{a\setminus i}\right) \right]$


Under these general settings the local dynamics of spin $i$ is described by a local Master Equation:

\begin{eqnarray}
\nonumber
  \dfrac{dP(\sigma_i)}{dt} = -\sum_{\sigma_{\partial i }} [ r_i(\sigma_i, \sigma_{ \partial i}) P(\sigma_i, \sigma_{ \partial i}) - \\ 
    r_i(-\sigma_i, \sigma_{\partial i }) P(-\sigma_i, \sigma_{ \partial i}) ]
\label{eqn:local_master_equation}
\end{eqnarray}

\noindent Equation (\ref{eqn:local_master_equation})  looks simpler than (\ref{eq:master-equation}), but  it is not a closed set of equations. To close it we need to resort to proper approximations. In this case we first write $P(\sigma_i,\sigma_{ \partial i })=\prod_{a \in \partial i} P(\sigma_{a \setminus i}|\sigma_i) P(\sigma_i)$, which is exact at equilibrium for trees and random graphs with large loops, and substitute it in (\ref{eqn:local_master_equation}). The goal in what follows is to find a proper approximation for the dynamics of the conditional probability $P(\sigma_{a \setminus i}|\sigma_i)$ that substituted back in (\ref{eqn:local_master_equation}) makes the calculation of $P(\sigma_i,t)$  straightforward.


We build our equations with the help of the theory of Random Point Processes \cite{vanKampen,Kipnis} where the dynamics of the spin variable $\sigma_i$ is encoded in a trajectory $\{X_i\}$ that is parametrized by the number of jumps $s_i$ of the corresponding spin in the studied
interval $[t_0,t]$ and the times $\{t^i_1, t^i_2, \dots, t^i_{s_i}\}$. Very generally for spins interacting through Hamiltonians defined on factor graphs the joint probability distribution $Q$ of these trajectories can be written as: $Q(X_{1}\ldots X_{N})=\prod_{i=1}^N \Phi_{i}(X_{i}\mid X_{\partial i})$ where

\begin{eqnarray}
\nonumber
\Phi_{i}(X_{i}\mid X_{\partial i})&=& \prod_{l=1}^{s_{i}} r_i(\sigma_{i}(t_{l}),\sigma_{\partial i}(t_{l}))\\
&\times&\exp[ -\int_{t_{0}}^{t}r_i(\sigma_{i}(\tau),\sigma_{\partial i}(\tau))d\tau].
\end{eqnarray}

For locally tree-like graphs this parametrization leads to a message-passing equation (see Supplementary Material), structurally identical
to the Belief Propagation equations used to approximate an equilibrium Gibbs-Boltzmann distribution\cite{mezard2009information}:

\begin{eqnarray}
  \nonumber
  \mu_{a\rightarrow (i)}(X_{ a\setminus i}|X_i) = \prod_{j\in a \setminus i} \sum_{\substack{X_{b} | b \in \partial j \setminus a}} \Phi_j(X_j|X_{ \partial j }) \\ 
  \times\prod_{b\in \partial j\setminus a} \mu_{b\rightarrow (j)}(X_{ b\setminus j}|X_j)
 \label{eqn:update_bethe}
\end{eqnarray}

Formally, equation (\ref{eqn:update_bethe}) defines a set of fixed point equations for the probabilities $\mu_{a\rightarrow (i)}(X_{ a\setminus i}|X_i)$ of the histories in the set $X_{ a\setminus i}$ in terms of equivalent objects in the neighborhood of the spins in $a\setminus i$, considering that the history $X_i$ is given.

Unfortunately, the histories, defined by the variables $X$ are cumbersome arguments to treat beyond formal statements, and actual solution of
(\ref{eqn:update_bethe}) is hopeless. One can however marginalize this quantity via $p(\sigma_{a \setminus i}| X_i,t) = \displaystyle \sum_{\substack{X_{ a\setminus i} \\ \sigma_{a\setminus i}(t)=\sigma_{a\setminus i}}} \mu_{a\rightarrow (i)}(X_{a \setminus i}|X_i)$, and then take the proper time derivative (see Supplementary Material) to obtain:

\begin{eqnarray}
\nonumber
 \lefteqn{\dot{p}(\sigma_{ a \setminus i}| X_i)=-\sum_{j\in a\setminus i} \sum_{\substack{\lbrace \sigma_{ b\setminus j }\rbrace \\ b \in \partial j \setminus a}} 
r_j^{+}  p(\lbrace \sigma_{ b\setminus j } \rbrace_{b \in \partial j \setminus a},\sigma_{ a \setminus i} \mid X_{i})}\\ 
 &&+ \sum_{j\in a\setminus i} \sum_{\substack{\lbrace \sigma_{b\setminus j }\rbrace \\  b \in \partial j \setminus a}} r_j^{-} p(\lbrace \sigma_{b \setminus j } \rbrace_{b \in \partial j \setminus a},F_j[\sigma_{a \setminus i} ]\mid X_{i})
\label{eq:Tree_exact_CME}
\end{eqnarray}

\noindent where $r_j^{+/(-)}= r_j((-)\sigma_j , \lbrace \sigma_{ b\setminus j } \rbrace_{b \in \partial j \setminus a}, \sigma_{ a \setminus j})$ and $\lbrace \sigma_{ b \setminus j}\rbrace$ with  $ b \in \partial j  \setminus a$ is the set of instantaneous variables that characterize the nodes neighboring $j$, except those in $a$.

Note that although eq. (\ref{eq:Tree_exact_CME}) is exact in tree-like graphs it still contains spin histories as conditional arguments. In principle, these histories could be taken as parameters to be tracked during the solution of  (\ref{eq:Tree_exact_CME}). However, it is convenient to go further assuming first that variables factorize around factor nodes: $p(\lbrace \sigma_{ b\setminus j } \rbrace_{b \in \partial j \setminus a} , \sigma_{ a \setminus i} \mid X_{i}) = p(\lbrace \sigma_{ b\setminus j } \rbrace_{b \in \partial j \setminus a} \mid \sigma_{ a \setminus i},  X_{i}) p ( \sigma_{ a \setminus i} \mid X_{i}) \approx \displaystyle \prod_{b \in \partial j \setminus a} p( \sigma_{ b\setminus j } \mid  \sigma_{ a \setminus i})p ( \sigma_{ a \setminus i} \mid X_{i})$ which is exact in tree-like graphs in equilibrium. Then, to close the system of equations it is enough to consider that locally the system has short time memory,  $p(\sigma|X_i) \sim p(\sigma|\sigma_i)$ which is equivalent to a Markov hypothesis.

With these approximations equation (\ref{eq:Tree_exact_CME}) transforms into a Master Equation but conditioned to neighboring (cavity) spins. We call this the Cavity Master Equation (CME):


\begin{eqnarray}
\nonumber
 \lefteqn{\dot{p}(\sigma_{ a \setminus i}| \sigma_i)=-\sum_{j\in a\setminus i} \sum_{\substack{\lbrace \sigma_{ b\setminus j }\rbrace \\ b \in \partial j \setminus a}} 
r_j^{+} \displaystyle \prod_{b \in \partial j \setminus a} p( \sigma_{ b\setminus j }\mid \sigma_{j})\;p(\sigma_{a \setminus i} \mid \sigma_{i})}\\
 &&+ \sum_{j\in  a\setminus i} \sum_{\substack{\lbrace \sigma_{ b\setminus j }\rbrace \\ b \in \partial j \setminus a}} 
 r_j^{-} \displaystyle \prod_{b \in \partial j \setminus a} p( \sigma_{b\setminus j }\mid -\sigma_{j})\;p(F_j[\sigma_{ a \setminus i}] \mid \sigma_{i})
\label{eq:CME}
\end{eqnarray}


Then,  to obtain the observables of the system one just solves the set of equations (\ref{eq:CME}), plugs the result in the local Master Equation (\ref{eqn:local_master_equation}) and integrate to obtain the local probabilities at the nodes of the network. This gives the joint probability distribution of all the variables for all times in one go,
the same outcome which would require averaging over many realizations using Kinetic Monte Carlo (KMC). Although we will not pursue this issue here, let us note that a similar approach could also, as for standard Belief Propagation, be used to describe the evolution of the marginal probability of a simply connected subset of nodes in the graph.



Let us now discuss in more detail some of the properties of the p-spin ferromagnetic model with $p=3$ in a random regular graph with $k = 3$. As discussed in \cite{Franz} it shows three different phases. For $T>T_{ms}=1.63$ it is paramagnetic. At $T_{ms}$ emerges a ferromagnetic metastable consistent with a spinodal transition. This ferromagnetic state becomes stable at $T_{fm}=1.21$. For lower temperatures $T < T_{K}=0.655$\cite{Montanari04} the system has a thermodynamic phase transition to a spin glass state. The origin of this spin glass states is the existence of a free-energy minimum with the same energy  of the ferromagnetic state, but with higher entropy. This minimum can be described as follows: when a spin is down, it puts an effective negative interaction on the other $k-1=2$ neighbors that will act as an anti-ferromagnetic pair. Thus, given the tree-like structure of the lattice, is is possible to minimize the energy with a mix of ferromagnetic and anti-ferromagnetic plaquettes. The previous configurations are energy minimum and their mean magnetization is zero. There are many different realizations of such configurations, and this leads to fluctuations in the system dynamics, whose interactions are not fully satisfied because of existing loops. Eventually the spins freeze giving origin to spin glass states that define the dynamics of the model for $T_d < 0.757$\cite{Montanari04}.

We know from previous results that the CME fails to predict KMC results deep into the SG phase and for intermediate time scales near second order phase transitions\cite{Aurell}.  Therefore we will here focus first our attention in the behavior of the method around the spinodal transition. We compare CME to KMC starting at $t=0$ from a totally ordered configuration with magnetization one. Then we observe the evolution of the average magnetization after a quench to a temperature $T$ and quantitatively compare the CME and KMC using the mean square difference of local magnetization's $\delta m(t)=\sqrt{\frac{1}{N}\sum_{i}(m^{CME}_i(t)-m^{KMC}_i(t))^2}$

\begin{figure}[!htb]

  \includegraphics[keepaspectratio=true,width=0.4\textwidth]{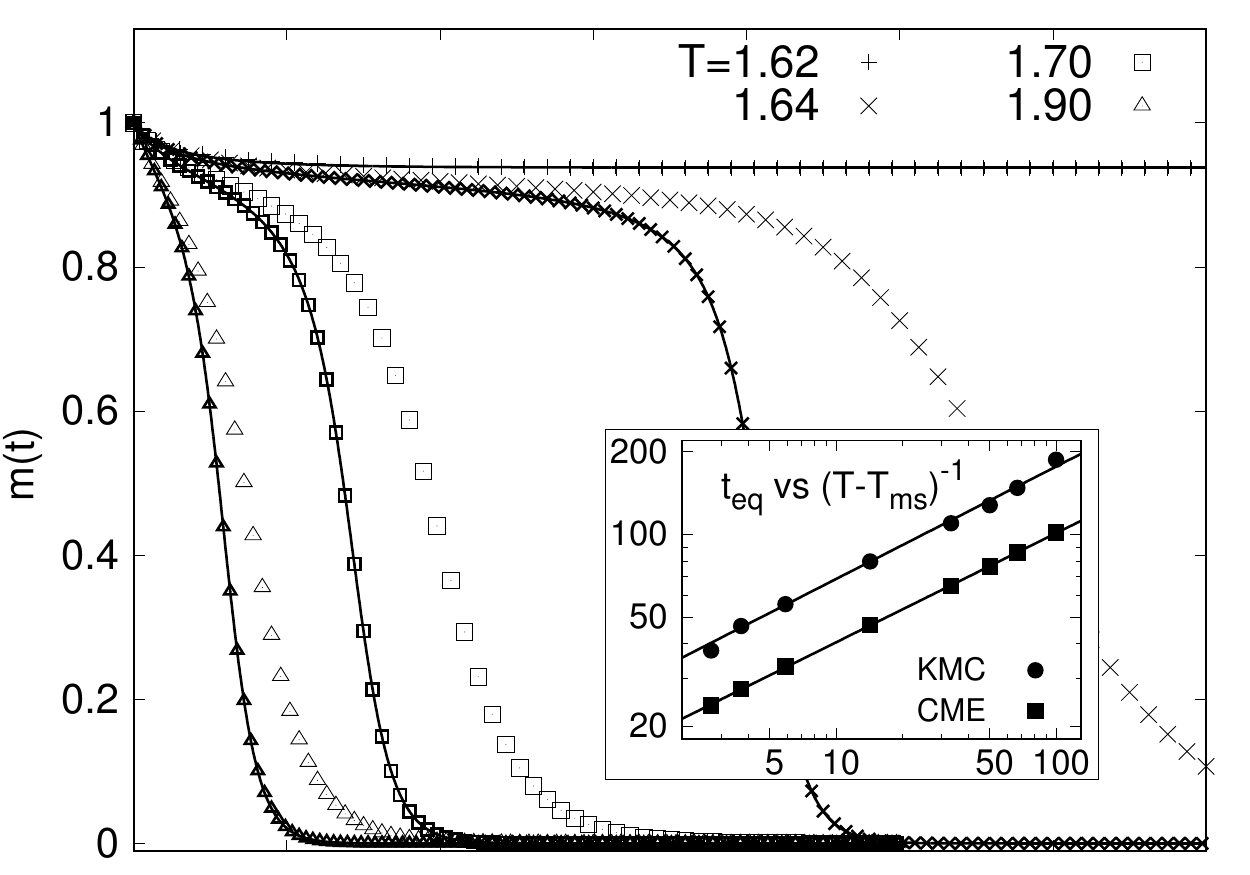}
  \includegraphics[keepaspectratio=true,width=0.4\textwidth]{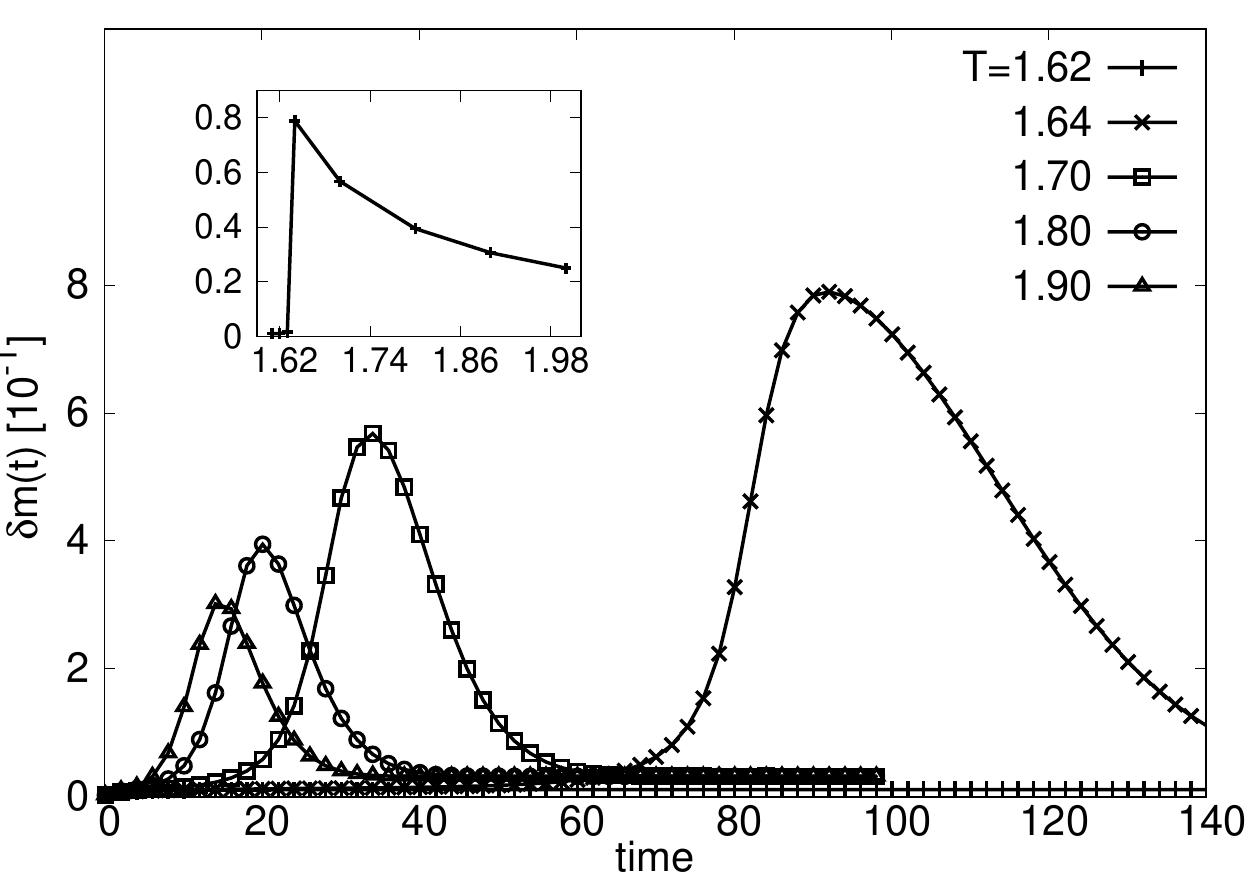}

  \caption{Top: Dynamics of the average magnetization using CME (lines and symbols) and KMC (symbols).  In the inset: Equilibration time near the spinodal transition temperature. Bottom: Dynamical behavior of the local error  between CME and KMC. System size: $N=99999$ spins. In the inset: Temperature dependence of the maximum local error. At time $t=0$ the system is fully magnetized, and then begins to interact with a heat bath at a given temperature.}
\label{fig:the_dynamics_CME}
\end{figure}




Results for both quantities are reported in panels of Fig.~\ref{fig:the_dynamics_CME}. In this numerical experiment, below the spinodal transition ($T_{ms}<1.63$), and deep in the high temperature region ($T>1.90$)  CME and KMC produce the same output. In the first case the system stays trapped in the magnetized state while in the later the system relax exponentially to a paramagnetic phase.  Only above and near the spinodal transition and at intermediate times do
the algorithms differ; the long-term results, however, are the same.
Note that even around the spinodal transition the errors ultimately become zero supporting the idea that the CME properly reproduces the long time behavior of the system.

\begin{figure}[!htb]
\includegraphics[keepaspectratio=true,width=0.4\textwidth]{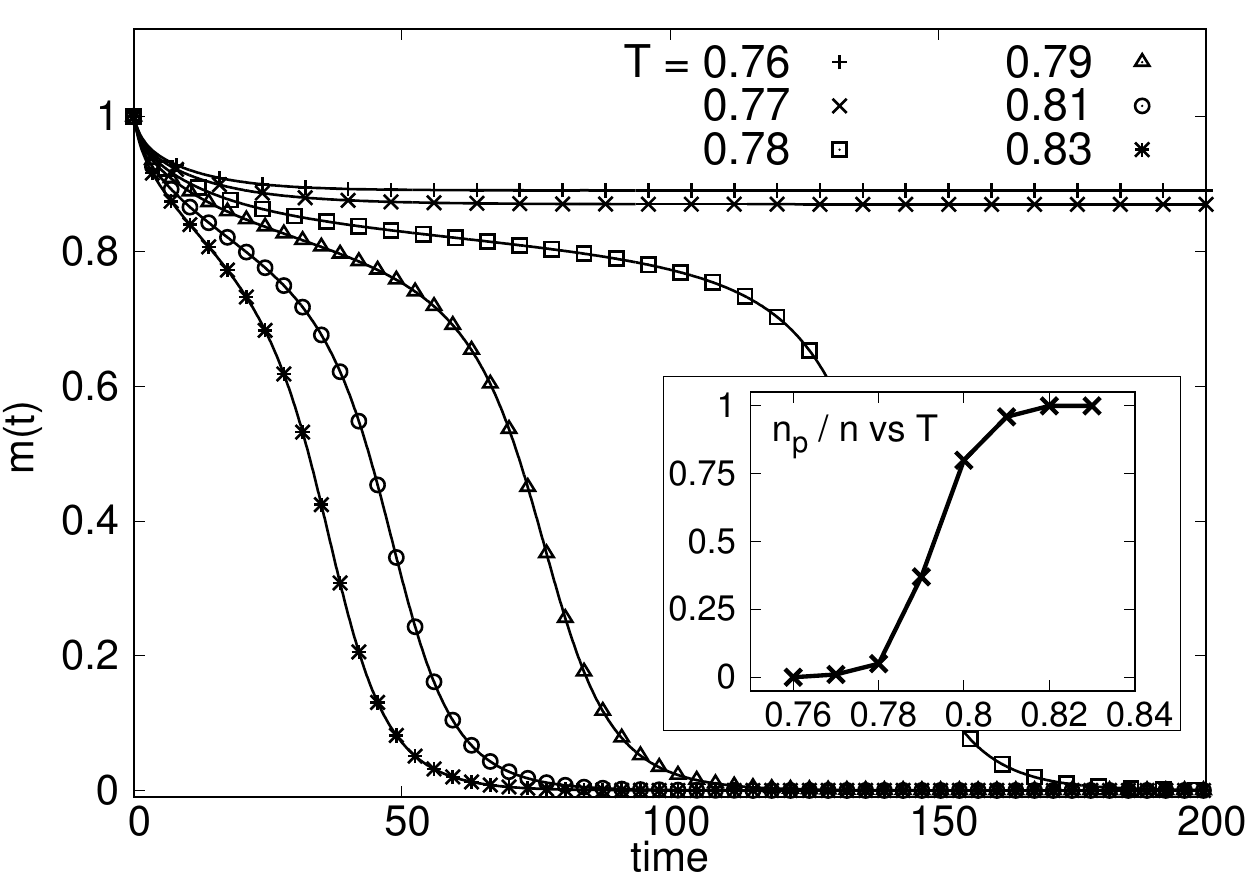}
\caption{Dynamical behavior of the planted model for $N=10002$ near $T_d$. In the plot are represented curves for a single planted graph. In the inset we show the fraction of the samples in which the magnetization decays to zero as a function of $T$.}
\label{fig:Plantedmodel}
\end{figure}

We also tested our CME in a planted model \cite{Krzakala2016} near the dynamical phase transition of the model $T_d$. The planted model is built fixing the ground state, all $\sigma_i=1$, and then choosing the value of the links in the graph following the rule: $P(J_a=1) = \frac{1+\tanh(\beta)}{2}$ and $P(J_a=-1) = \frac{1-\tanh(\beta)}{2}$. The results of the evolution of the magnetization in this graph are shown in Fig.~\ref{fig:Plantedmodel}. In this case our dynamics evolves toward the correct paramagnetic state for $T>.78$ a temperature that is close to but slightly above the expected $T_d=0.757$.  However, one most notice, that these results are strongly affected by sample to sample fluctuations, as the inset in Figure .~\ref{fig:Plantedmodel} shows.

\begin{figure}[!htb]
  \includegraphics[keepaspectratio=true,width=0.4\textwidth]{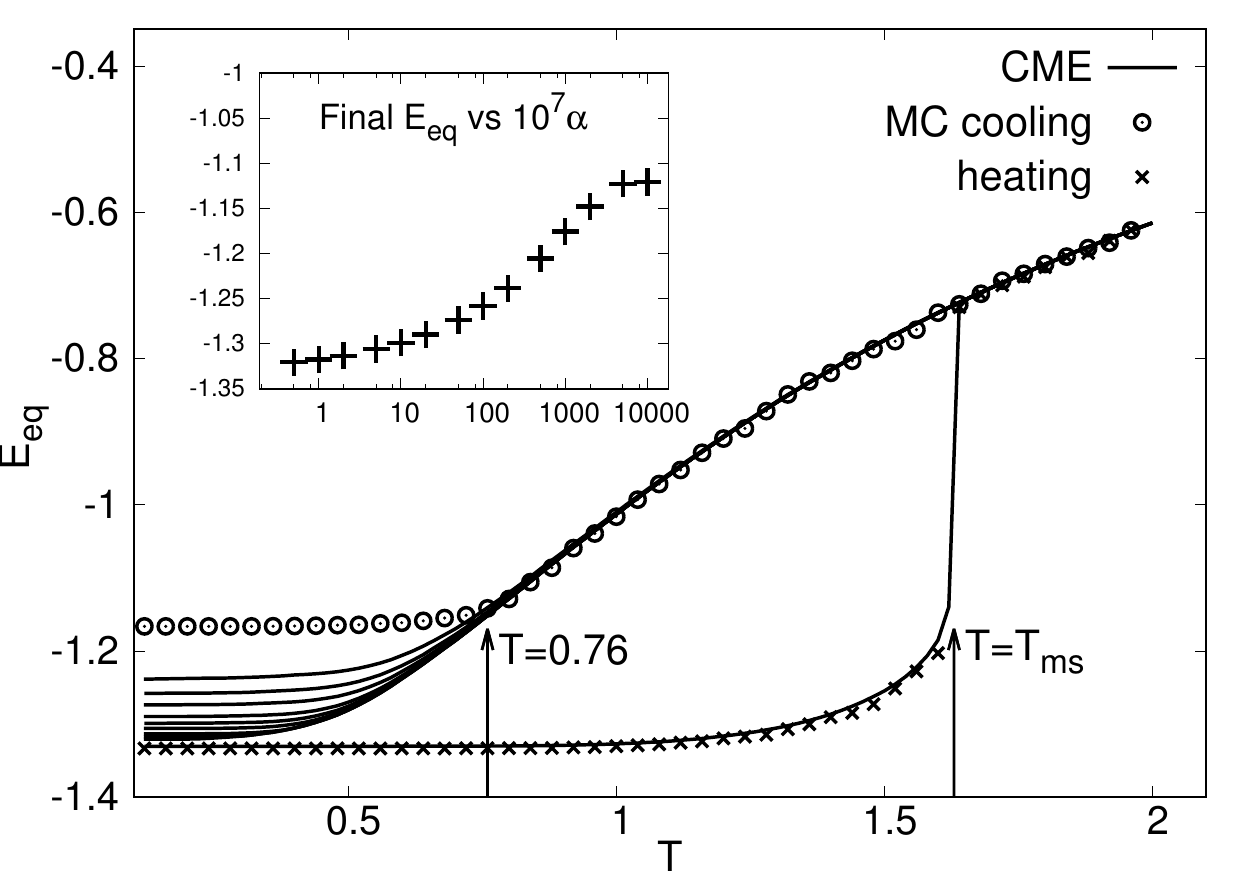}
 
\caption{Comparison between the quasi-equilibrium evolution of the CME (continuous lines) and KMC (dots). The CME was applied to a $N=1002$ system, and KMC to a $N=10002$ system. The temperature step is in both cases $\Delta T = 0.02$. Both approaches reproduce the existence of $T_{ms} \approx 1.62$. Below $T_d$ both dynamics differentiate}
  \label{fig:equilibrium_comparison_MC_CME}
\end{figure}

Motivated by the long time behavior of the CME we explored the behavior of the observables of the system in an adiabatic protocol. Although the set of differential equations (\ref{eq:CME}) was written with the aim to study dynamical processes one can equally well let the integration evolve in time and define the equilibrium states when the cavity probabilities reach a fixed point of the dynamics. This idea is used here to reproduce adiabatic heating and cooling experiments in the model. In Fig.~\ref{fig:equilibrium_comparison_MC_CME} we show the result of these experiments and compare them with KMC. The heating process reflects quite well the existence of $T_{ms}$. Starting from an homogeneous ferromagnetic configuration at low $T$, the temperature is increased in small steps. Both CME and KMC remain in the ferromagnetic state until it disappears beyond the spinodal temperature and then jumps to the paramagnetic state. 

On the other hand, if we start cooling the system from a paramagnetic state at high temperatures (Figure (\ref{fig:equilibrium_comparison_MC_CME})), the CME and KMC will coincide only down to the temperature $T_d\approx 0.757$ for which the spin glass phase dynamically traps the stochastic simulation \cite{Franz}. Below this temperature the behavior of the CME depends on the criterion adopted to define the convergence of the algorithm. Our criterion is to stop whenever for each plaquette $\mid p(\sigma_k,\sigma_j \mid \sigma_i, t+ dt) - p(\sigma_k,\sigma_j \mid \sigma_i, t+)\mid \leq \alpha$, where $\alpha$ is a parameter controlling the time that the system spends at a given temperature. The smaller the $\alpha$ the larger the time we gave the system to equilibrate. Remarkably, if $\alpha$ is small enough the probability measure of the CME splits evenly among ordered states with antiferromagnetic and ferromagnetic plaquettes of the same energy.  Moreover, not only the total but also the local magnetizations obtained in this temperature region are zero. This suggests that the quasi-dynamics given by the CME deep into the glassy phase may be useful to explore the low energy structure of the system at low temperatures. 

\begin{figure}[!htb]
  \includegraphics[keepaspectratio=true,width=0.4\textwidth]{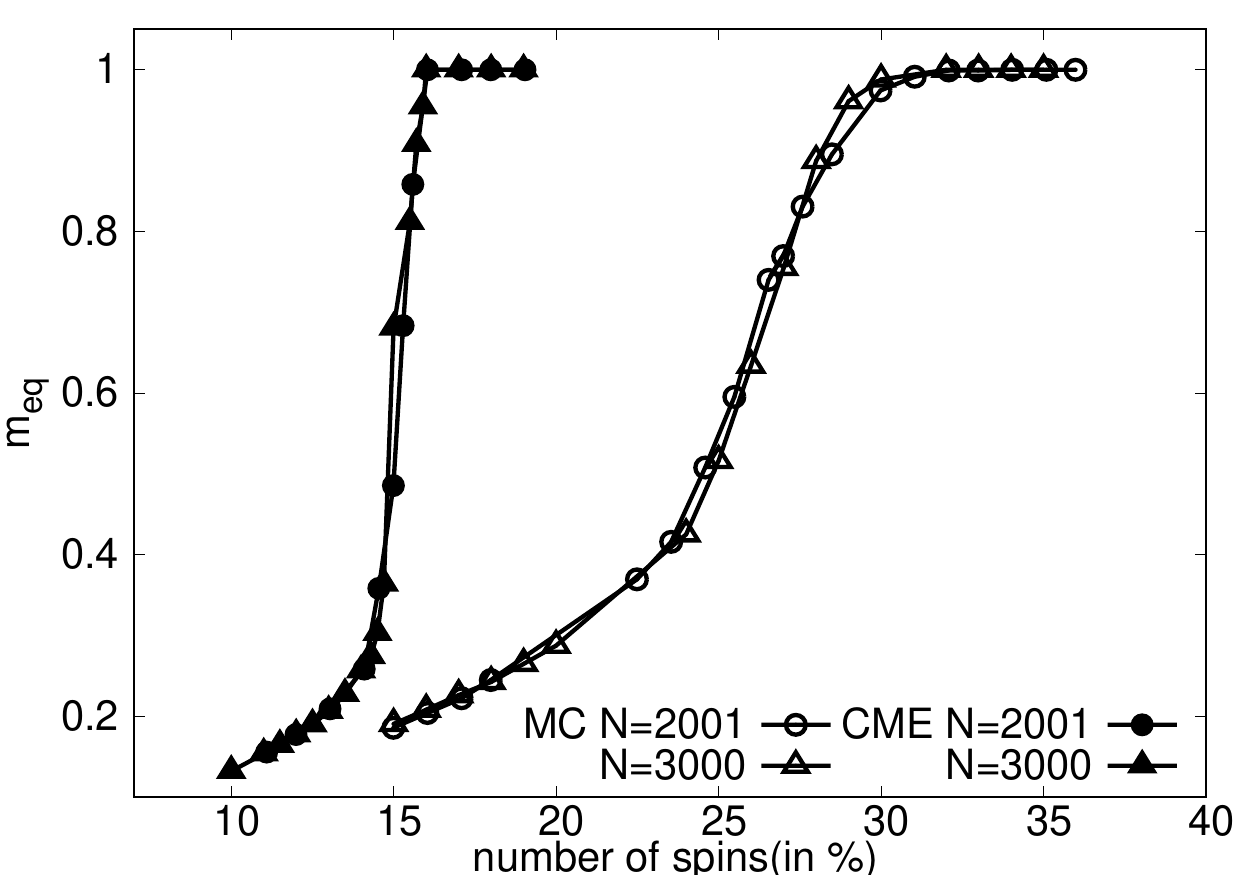}
  \caption{Final magnetization of a system as a function of the number of fixed spins at $T=0.12$ following the dynamics of the CME $N=2001$ and KMC $N=3000$.}
  \label{fig:cluster}
\end{figure}

This low temperature structure is indeed one of the most interesting features of the model, since despite the clear presence of a crystalline state it can not be reached by any known local dynamical rule.  In that scenario, we decided to fix a fraction of the spins using strong local fields pointing in the same direction.
We then inquired how large should be the fraction of spins aligned to obtain the correct crystalline state after a quench to a very low temperature.
 The results are shown in Figure \ref{fig:cluster} where we compare the results of the CME and KMC. It is evident that while the KMC dynamics converges to the proper equilibrium state when almost $30\%$ of the spin are oriented in the ferromagnetic state, the pseudo-dynamics of the CME at low temperature recognize the ground state of the model with only a $15\%$ of the spins correctly oriented. The smaller number of spins needed to drive the dynamics of the model into the ground state of the system for the CME suggests that it can be used as proper proxy for similar problems in Combinatorial Optimization.

In summary, we have presented a formal approach to study the continuous dynamics of a discrete system with multiple-particle interactions. We ilustrate the power of this approach studying the dynamical behavior of the $p$-spin dilute ferromagnet. We have shown that the Cavity Master Equation reproduces the long time behavior of Kinetic Monte Carlo simulations in a wide temperature range. CVM also reproduces exactly the spinodal Temperature $T_{ms}$ and is a very good proxy for the exact value of the dynamical temperature $T_d$. Below $T_d$, where the spin glass states dominate the dynamics the CME fails to predict the KMC results. However, this allows a deeper exploration of the structure of the glassy phase and in particular to find the ground state of the system by fixing a small number of spins.

\textbf{Acknowledgments:} This project has received funding from the European Union’s Horizon 2020 research and innovation programme MSCA-RISE-2016 under grant agreement No. 734439 INFERNET. It has been partially supported by the STINT (Sweden) project "Enhancing cooperation opportunities with Havana University through the Erasmus Programme actions", by an Erasmus+ International Credit Mobility to KTH (EU), and by the COIN Centre of Excellence,'of the Academy of Finland (grant no. 251170. 



\bibliographystyle{unsrt}
\bibliography{ref_cvme}

\onecolumngrid

\section{Supplementary Material}

\subsection{Message Passing Equation}
A possible parametric expression for the joint probability of the spin trajectories is:

\begin{equation}
Q(X_{1}\ldots X_{N})=\prod_{i=1}^N \Phi_{i}(X_{i}\mid X_{\partial i})
\label{eqn:SMhistorias}
\end{equation}
where
\begin{equation}
\Phi_{i}(X_{i}\mid X_{\partial i})= \prod_{l=1}^{s_{i}} r_i(\sigma_{i}(t_{l}),\sigma_{\partial i}(t_{l}))
\times\exp[ -\int_{t_{0}}^{t}r_i(\sigma_{i}(\tau),\sigma_{\partial i}(\tau))d\tau]
\end{equation}

\normalsize

For a tree-like structure the expression (\ref{eqn:SMhistorias}) can be written as an expansion that takes a factor node as a starting point. In a mathematical language:

\begin{equation}
Q(X_{1},...,X_{N})=\Phi_{i}(X_{i}\mid X_{ \partial i})\prod_{ a \subset \partial i} M_{a\;i}\left(X_{a\setminus i},\{X\}_{a,i}\mid X_{i}\right)
\label{eqn:SM3}
\end{equation}
\noindent where $M_{a\;i}$ is:

\begin{equation}
  M_{a\;i}\left(X_{a\setminus i},\{X\}_{a,i}\mid X_{i}\right)=\prod_{j\in a\setminus i} \Phi_j(X_j|X_{ \partial j }) \prod_{b \in \partial j\setminus a} \left[\prod_{m\in b\setminus j} \Phi_m(X_m|X_{ \partial m }) \ldots \right]
\label{eqn:SM}
\end{equation}

Now, marginalizing conveniently:

\begin{equation}
    Q(X_i,X_{ \partial i }) = \Phi_i(X_i|X_{ \partial i }) \prod_{ a \subset \partial i} \mu_{a\rightarrow (i)}(X_{a\setminus i}|X_i)
     \label{eqn:SMmarginal_prob_dist_mess}
\end{equation}
where:
\begin{equation}
   \mu_{a\rightarrow (i)}(X_{a\setminus i}|X_i) = \sum_{\{X\}_{a,i}} M_{a\;i}\left(X_{a\setminus i},\{X\}_{a,i}\mid X_{i}\right)
\end{equation}
can be interpreted as probabilities of the histories in $X_{a \setminus i}$ given a fixed $X_i$, or in other words, in a cavity where $X_i$ is the history of spin $i$.

On the other hand, the nodes connected to a factor node can also be taken as starting point of a new expansion of (\ref{eqn:SM3}). Again, after marginalization:
\begin{equation}
    Q(X_i,X_{ \partial i }) =  \sum_{\substack{\lbrace X_{b} \rbrace \\ b \in \partial j \setminus a,\\ \forall j \in a \setminus i}} \prod_{j'\in a} \Phi_{j'}(X_{j'}|X_{ \partial j' }) \prod_{b\in \partial j'\setminus a} \mu_{b\rightarrow (j')}(X_{b\setminus j'}|X_{j'})
    \label{eqn:SMpair_prob_dist_mess2}
\end{equation}

and putting (\ref{eqn:SMmarginal_prob_dist_mess}) and (\ref{eqn:SMpair_prob_dist_mess2}) together gives an equation for $\mu$:
\begin{equation}
 \mu_{a\rightarrow (i)}(X_{a\setminus i}|X_i) = \sum_{\substack{\lbrace X_{b} \rbrace \\ b \in \partial j \setminus a,\\ \forall j \in a \setminus i}} \prod_{j'\in a \setminus i} \Phi_{j'}(X_{j'}|X_{ \partial j' }) \prod_{b\in \partial j'\setminus a} \mu_{b\rightarrow (j')}(X_{b\setminus j'}|X_{j'})
 \label{eqn:update_bethe_0}
\end{equation}

In order to seek some simplicity in (\ref{eqn:update_bethe_0}), one can interchange the product and the sum:
\begin{equation}
 \mu_{a\rightarrow (i)}(X_{a\setminus i}|X_i) = \prod_{j\in a \setminus i} \sum_{\substack{\lbrace X_{b} \rbrace \\ b \in \partial j \setminus a}} \Phi_j(X_j|X_{ \partial j }) \prod_{b\in \partial j\setminus a} \mu_{b\rightarrow (j)}(X_{b\setminus j}|X_j)
 \label{eqn:update_bethe_1}
\end{equation}

Note that equation (\ref{eqn:update_bethe_1}) gives the probabilities of the histories in any set $X_{a\setminus i}$ when $X_i$ is fixed in terms of similar objects that depends of all the neighbors of the nodes in $a\setminus i$. This derivation is exact only in tree-like graphs and good in Bethe lattices (with long loops).

\subsection{Time derivative}

The marginalization of $\mu_{a\rightarrow (i)}(X_{a\setminus i}|X_i)$ is:

\begin{equation}
 p(\sigma_{a \setminus i}| X_i,t) = \sum_{\substack{X_{a\setminus i} \\ \sigma_{a\setminus i}(t)=\sigma_{a\setminus i}}} \mu_{a\rightarrow (i)}(X_{a\setminus i}|X_i)
\end{equation}

\noindent and the derivative:

\begin{equation}
 \dfrac{dp(\sigma_{a \setminus i}| X_i,t)}{dt}= 
 \lim_{\Delta t \rightarrow 0}\dfrac{1}{\Delta t}\left[ p(\sigma_{a \setminus i}| X_i,t+\Delta t) - p(\sigma_{a \setminus i}| X_i,t)\right]
\end{equation}

Then,  to ease the notation from it will be written $\Delta t$ in place of $t + \Delta t$ and $0$ in place of $t$. So the previous formula would be:
\begin{equation}
 \dfrac{dp(\sigma_{a \setminus i}| X_i,t)}{dt}= 
 \lim_{\Delta t \rightarrow 0}\dfrac{1}{\Delta t}\left[ p(\sigma_{a \setminus i}| X_i,\Delta t) - p(\sigma_{a \setminus i}| X_i,0)\right]
 \label{eq:SMincremental_quotient}
\end{equation}

The first term inside the limit is the important one:

\begin{equation}
p(\sigma_{a \setminus i}| X_i,\Delta t) = \sum_{\substack{X_{a\setminus i}(\Delta t) \\ \sigma_{a\setminus i}(\Delta t)=\sigma_{a\setminus i}}} 
 \mu_{a\rightarrow (i)}(X_{a\setminus i}(\Delta t)|X_i(\Delta t))
 \label{eq:SMmarginal_p_delta_t}
\end{equation}

With a precision of $o(\Delta t)$, the trace in \eqref{eq:SMmarginal_p_delta_t} can be divided in two terms $A$ and $B$ such that:

\begin{equation}
p(\sigma_{a \setminus i}| X_i,\Delta t) = A + B + o(\Delta t) 
 \label{eq:SMmarginal_p_delta_t_1}
\end{equation}
\noindent where
\begin{equation}
 A = \sum_{\substack{X_{a\setminus i}(\Delta t) \\ \sigma_{a\setminus i}(0)=\sigma_{a\setminus i}}} 
 \mu_{a\rightarrow (i)}(X_{a\setminus i}(\Delta t)|X_i(\Delta t))
\end{equation}
\noindent and
\begin{equation}
 B = \sum_{j \in a\setminus i } \sum_{\substack{X_{a\setminus i}(\Delta t) \\ \sigma_{a\setminus i}(0)=F_j[\sigma_{a\setminus i}]}} 
 \mu_{a\rightarrow (i)}(X_{a\setminus i}(\Delta t)|X_i(\Delta t))
\end{equation}

Expanding A and B also to order $\Delta t$ and putting it together into the limit:

\begin{eqnarray}
\nonumber
 \lefteqn{\dfrac{dp(\sigma_{a \setminus i}| X_i,0)}{dt}=-\sum_{j\in a\setminus i} \sum_{\substack{X_{a\setminus i} \\ \sigma_{a\setminus i}(t)=\sigma_{a\setminus i}}} 
 \lambda_j (X_j,X_{a\setminus i,j},X_i) \mu_{a\rightarrow (i)}(X_{a\setminus i}|X_i)}\\
 &&+ \sum_{j\in a\setminus i} \sum_{\substack{X_{a\setminus i} \\ \sigma_{a\setminus i}(t)=F_j[\sigma_{a\setminus i}]}} 
 \lambda_j (X_j,X_{a\setminus i,j},X_i) \mu_{a\rightarrow (i)}(X_{a\setminus i}|X_i)
\label{eq:SMexact_cavity_lambdas}
\end{eqnarray}

Now it's needed an expression for $ \lambda_j (X_j,X_{a\setminus i,j},X
_i) \mu_{a\rightarrow (i)}(X_{a\setminus i}|X_i)$. In order to find it
, one can differentiate (\ref{eqn:SMupdate_bethe}) but with some care, because an assumption is required: the fact that the only nodes that can change its state in the $[ t, t+ \Delta t ]$ time interval are those in $a$ implies that the traces for the other nodes should be done to time t. Inserting the result into (\ref{eq:SMexact_cavity_lambdas}) and explicitly doing some required marginalizations gives:

\begin{eqnarray}
\nonumber
 \lefteqn{\dot{p}(\sigma_{a \setminus i}| X_i,0)=-\sum_{j\in a\setminus i} \sum_{\substack{\lbrace \sigma_{b\setminus j }\rbrace \\ b \in \partial j \setminus a}} 
r_j^{+} p(\lbrace \sigma_{b\setminus j } \rbrace_{b \in \partial j \setminus a},\sigma_{a \setminus i} \mid X_{i})}\\
 &&+ \sum_{j\in a\setminus i} \sum_{\substack{\lbrace \sigma_{b\setminus j }\rbrace \\ b \in \partial j \setminus a}} 
 r_j^{-} p(\lbrace \sigma_{b\setminus j } \rbrace_{b \in \partial j \setminus a},F_j[\sigma_{a \setminus i} ]\mid X_{i})
\label{eq:SMTree_exact_CME}
\end{eqnarray}

\noindent where $r_j^{+}= r_j(\sigma_j , \lbrace \sigma_{b\setminus j } \rbrace_{b \in \partial j \setminus a}, \sigma_{a \setminus j})$, $r_j^{-} = r_j(-\sigma_j , \lbrace \sigma_{b\setminus j } \rbrace_{b \in \partial j \setminus a}, \sigma_{a \setminus j})$.


\end{document}